\documentclass {jpconf}
\usepackage{amsmath}
\usepackage[pdftex]{graphicx}

\begin{document}
\title{ Anisotropy results from 193 GeV U+U collisions at RHIC}
\author{ Yadav Pandit (for the STAR Collaboration) }
\address{Department of Physics, University of Illinois at Chicago, USA}
\ead{ypandit@uic.edu}

\begin{abstract}
We report the measurement of azimuthal  anisotropy $v_{n}$ for \textit {n} = 1-5 as a function of transverse momentum $p_{T}$ and centrality in U+U collisions at $\sqrt{s_{NN}}$ = 193 GeV, recorded with the STAR detector at RHIC.  We also present results on $v_{2}$  and  azimuthal correlator related to charge separation across the reaction plane due to local parity violation(LPV)  in the ultra-central collisions. 
\end{abstract}

 \section{Introduction}
 The azimuthal anisotropy of particle production is commonly used in high energy nuclear  collisions to study the early evolution of the expanding system~\cite{methodPaper}. The prolate shape of the  Uranium  nuclei makes U+U collision specially interesting since it opens up a possibility  to produce more extreme conditions of excited matter at higher density and/or greater volume than  is possible using spherical nuclei like gold or lead at the same incident energy~\cite{UUSim}.  Central uranium-uranium collisions can have large elliptic deformation with �body on body� collisions, while for �tip on tip� collisions negligible deformation is expected. So U+U collisions may offer an opportunity to explore wider range of initial eccentricities. Study of flow harmonics with two different event selections, enhancing the contributions from one or the other configuration, will provide a new handle on the initial state effects, where large uncertainties in theoretical calculations still exist. 
 
\section{Analysis Details}
 Data reported in this proceedings were collected for U+U collisions at $\sqrt{s_{NN}}$ = 193 GeV with a minimum bias trigger  and the ultra central events  which were taken with a dedicated central trigger based on Zero Degree Calorimeter(ZDC)  with the STAR detector at RHIC. The Time Projection Chamber (TPC)~\cite{startpc} is the primary tracking detector at STAR. Event  used in this analysis are required to have the primary vertex  with in 30 cm along the beam direction and 2 cm in the  transverse direction from the center of the beam pipe.  The  event plane method and two particle cumulant method are used in the analysis.  In the event plane method, the event plane vector is reconstructed from tracks with transverse momentum ($p_{T}$) up to 2 GeV/$c$ and the subevents were separated by  $\eta$   gap of 0.2 units. Results are corrected for detector inefficiencies in both methods.    
\section{Results} 

\subsection{First and higher harmonics }

  \begin{figure}[h]
\center
\includegraphics[width=30pc]{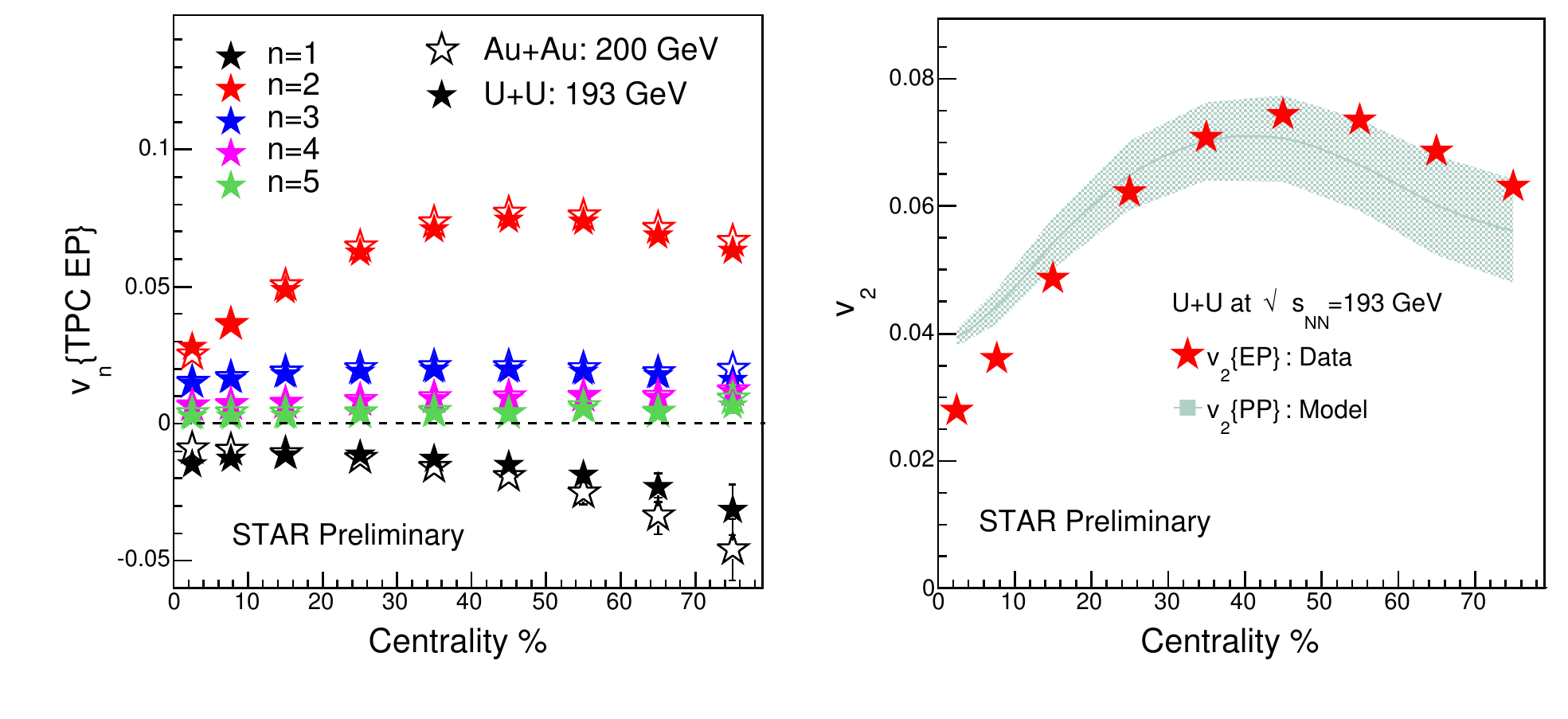}
\caption{ $v_{n}$ as a function of centrality on left and  $v_{2}$ as a function of  centrality compared with a model prediction.} 

\label{fig1}
\end{figure}

          Figure ~\ref{fig1} left panel  presents the  $v_{n}$ integrated in transverse momentum $p_{T}$ ($0.15 < p_{T}< 2 GeV/c$) and pseudorapidity $\eta$ ($|\eta|<1.0$) as a function of centrality.  Except for $v_{2}$, we observe very weak centrality dependence of all other harmonic coefficients which provides  possible hint that geometry fluctuations do not  strongly depend on centrality.  For comparison we also show the corresponding result for Au+Au collisions at 200 GeV.   On the right panel elliptic flow $v_{2}$  at 193 GeV compared with a model prediction based on Glauber model at 200 GeV~\cite{hiroshi}.  For central collisions, experimental data are lower  than the prediction.  

  \begin{figure}[h]
\center
\includegraphics[width=30pc]{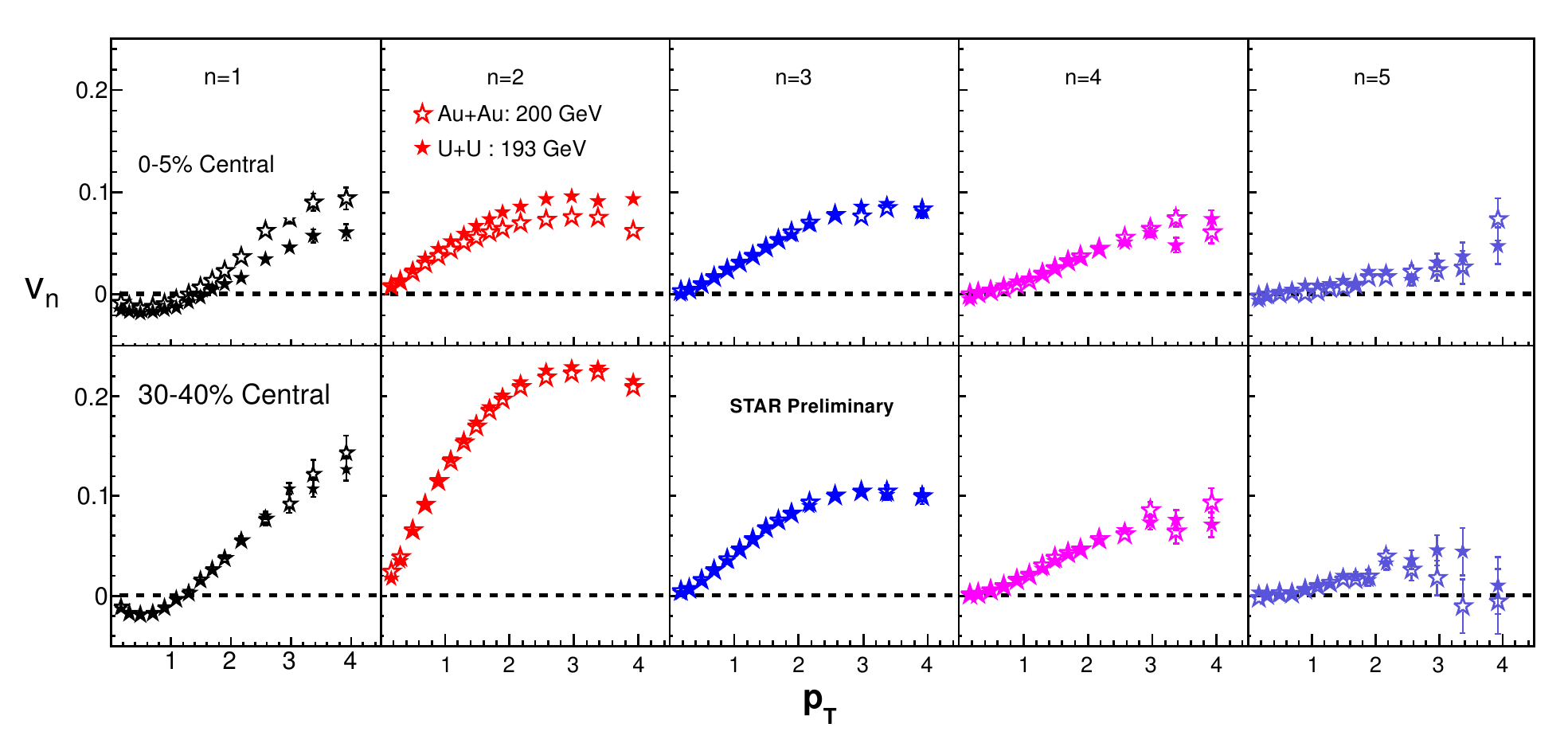}
\caption{$v_{n}(p_{T})$ measurement at 0-5\% central in upper panels and 30-40\% central collisions in the lower panels for U+U collisions at 193 GeV (solid stars) and for Au+Au collisions at 200 GeV(open stars).}
\label{fig2}
\end{figure}
\begin{figure}[h]
\center
\includegraphics[width=30pc]{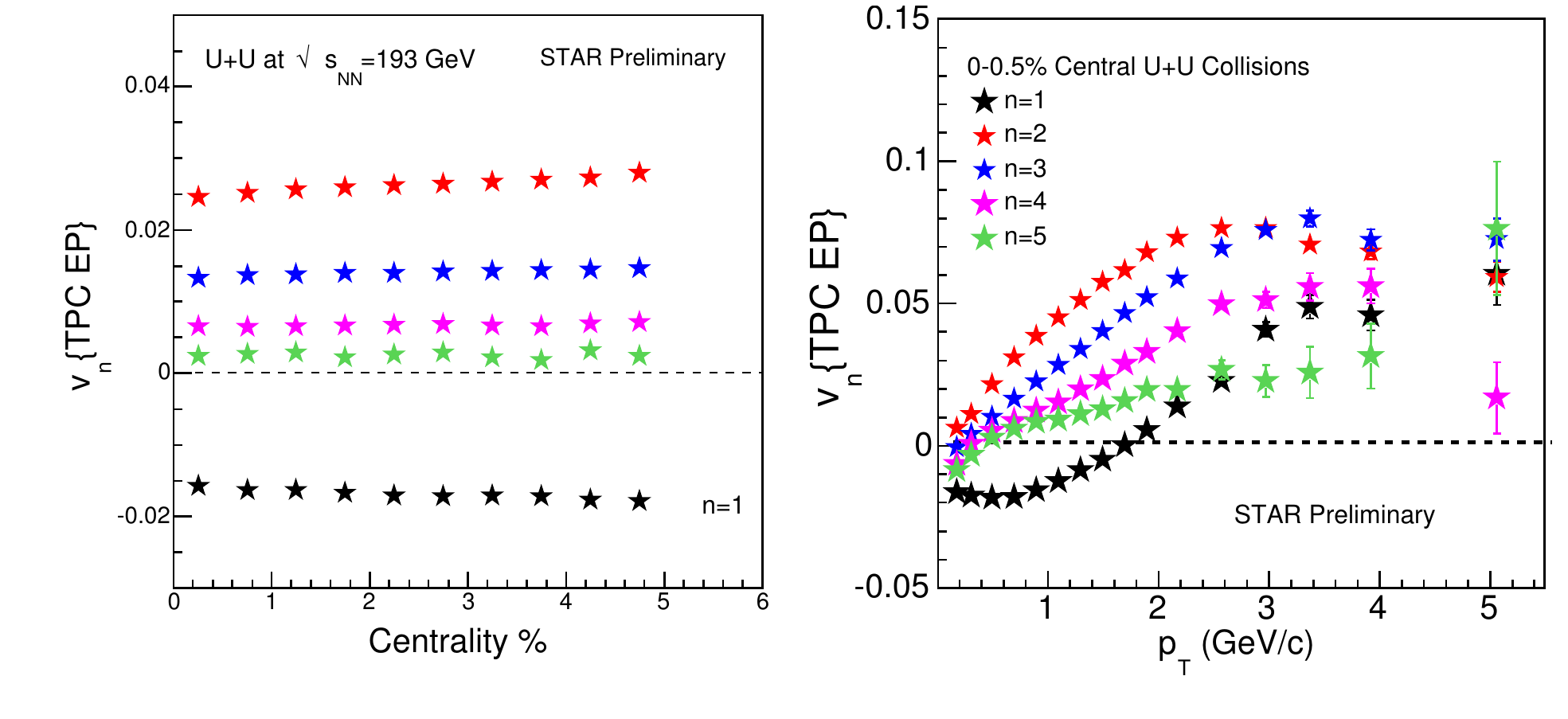}
\caption{$v_{n}$ as a function of centrality at 0-5\% central collisions on left and  $v_{n}$ as a function of $p_{T}$ at ultra central(0-0.5\%) collisions on right }
\label{fig3}
\end{figure}

        Fig.~\ref{fig2} shows  all $v_{n}$ measurements at different centralities  as a function of transverse momentum $p_{T}$. Also shown are the same measurement  from Au+Au collisions at 200 GeV for the comparisons. We observe the difference in $v_{n}$ for  \textit{n} = 1 and \textit{n} = 2 at 0-5\% central between U+U and Au+Au collisions. This may hint  to  initial overlap geometry difference in the central collisions between two systems Au+Au and U+U collisions. The difference diminishes in higher harmonics and more peripheral collisions.        
                  We also report measurement of $v_{n}$ for 0-5\% central U+U collisions taken with a dedicated central trigger. We subdivide 0-5\% central bin into 10 smaller centrality bins upto most central 0-0.5\% centrality.   In Fig.~\ref{fig3} left panel, $v_{n}$ as a function of centrality for ultra central collisions is shown. Other than second harmonic coefficient, the $v_{n}$  do not change in this centrality range. We observe small change for $v_{2}$, however we could not observe a knee structure predicted in some of the model calculations~\cite{Knee}.  Fig.~\ref{fig3} right panel shows  $v_{n}$ as a function of transverse momentum for the most central(0--0.5\%)  collisions which shows all harmonics coefficients have comparable magnitude in intermediate transverse momentum. 

\subsection{Fully overlap Collisions}
  \begin{figure}[h]
\center
\includegraphics[width=30pc]{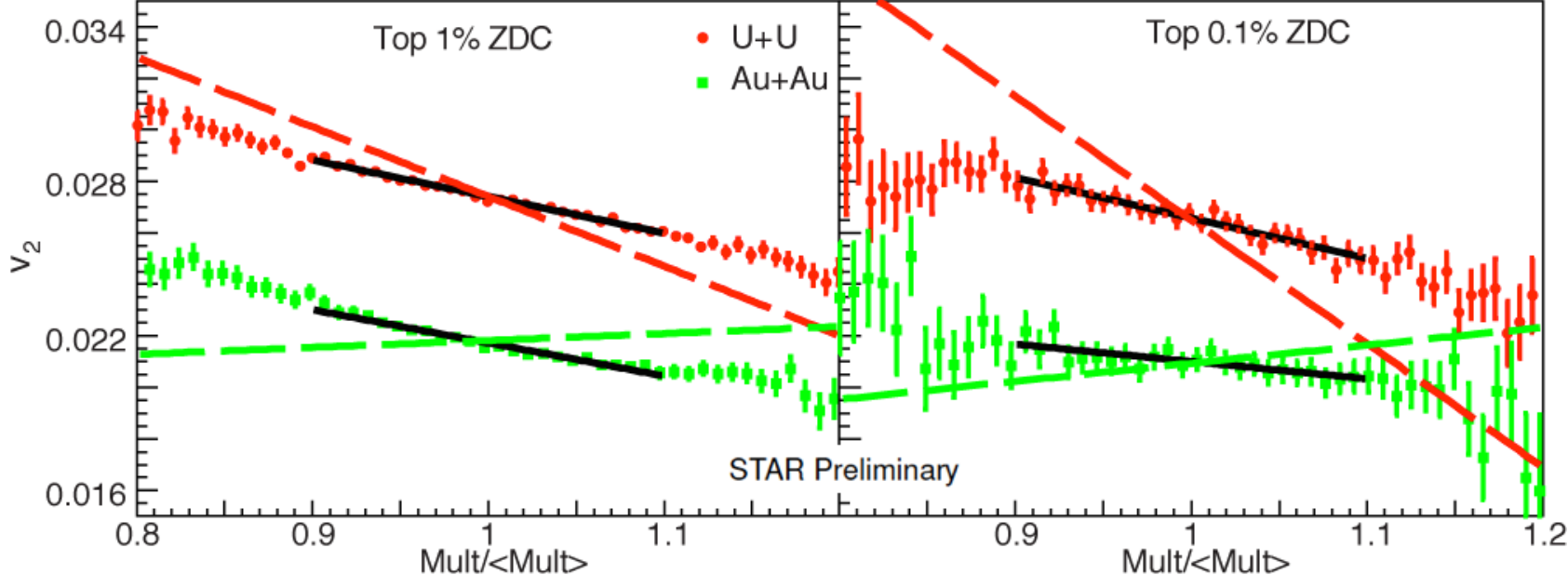}
\caption{The elliptic flow $v_{2}$ of all charged particles as a function of the normalized multiplicity. The left panel shows the results for top 1\% ZDC central events, while the right panel shows the results for top 0.1\% ZDC central event. A linear fit between normalized multiplicity 0.9 to 1.1 is applied to extract the slope parameter.The dash lines represent Glauber simulation slopes calculated via eccentricity }
\label{fig4}
\end{figure}
The Zero Degree Calorimeters (ZDCs) were used to select fully overlap event based on energy deposition by  spectator neutrons.  Figure 4 shows the $v_{2}$ of all charged particles as a function of the normalized multiplicity in 1\% ZDC centrality on left and 0.1\% centrality on right~\cite{Hui} . Since the deformation of gold nuclei is small, in case of fully overlapping collisions, we do not expect  correlations between multiplicity and $v_{2}$. However, we observe a strong negative slope in both Au+Au and U+U collisions for 1\% central data.  In order to see the effects from initial geometry (body on body vs. tip on tip), one needs to reduce the effects from impact parameters and select further on fully overlap collisions.  On 0.1\% selected events the slope magnitude in Au+Au  collisions becomes smaller  and  the slope for U+U become steeper as expected.  The Glauber model predicts a steeper slope for U+U collisions and a positive slope for Au+Au due to its oblate shape and dose not describe the data well.
 \subsection{Local Parity Violation}
 
 In heavy ion collisions  a strong magnetic field is produced because of  energetic spectator protons.  The interplay between the magnetic field and the quark-gluon matter created in the collisions may have observable effects~\cite{Dima}.  To study the effect of  charge separation due to the Chiral Magnetic Effect, a three-point mixed harmonics azimuthal correlator was proposed~\cite{Sergei2}:
\begin{equation}
\gamma = \langle \cos(\phi_{\alpha} + \phi_{\beta} - 2\psi_{\rm RP}) \rangle,
\label{eq:eq3}
\end{equation}
where $\alpha$ and $\beta$ denote the particle type: $\alpha$, $\beta = +$, $-$.

The observable $\gamma$ is {$\cal P$}-even, but sensitive to the fluctuation of charge separation. The left panel of Fig. 5 shows $\gamma$ as a function of centrality for U+U collisions at 193 GeV~\cite{gangWang}, in comparison with Au+Au collisions at 200 GeV  measured by STAR collaboration~\cite{STAR_LPV1}. The opposite sign correlations in U+U are still higher than the same sign, with $\gamma_{SS} $ consistent with those in Au+Au and $\gamma_{OS} $ slightly lower than those in Au+Au, showing the clear difference between the opposite sign and the same sign correlations, qualitatively consistent with the picture of CME and LPV.   
   
\begin{figure}[h]
\center
\includegraphics[width=30pc]{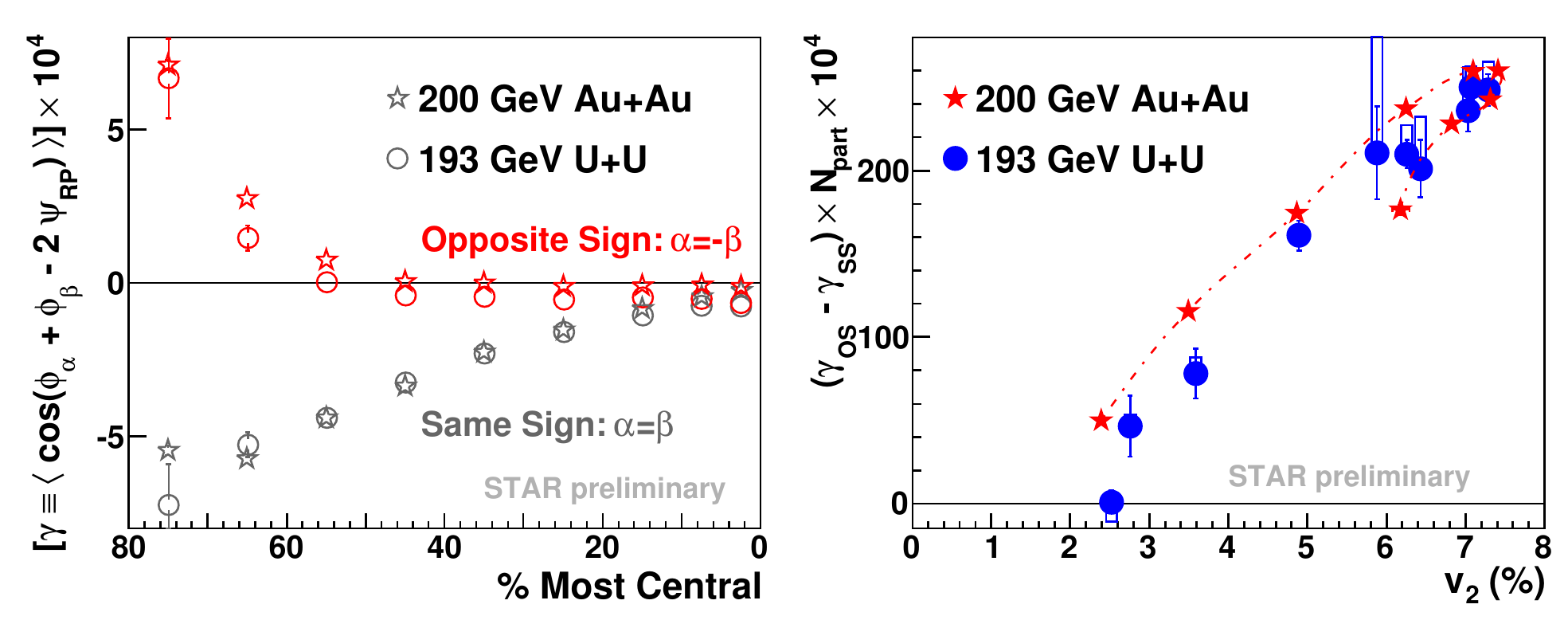}
\caption{(Color online) Comparison between Au+Au collisions at 200 GeV and U+U collisions at 193 GeV. (Left) $\gamma$ as a function of centrality. (Right) $(\gamma_{\rm OS}-\gamma_{\rm SS}) \times N_{part}$ vs $v_{2}$.The error bars are statistical only. The open box represents the systematic uncertainty due to the tracking capability under the high} 

\label{fig5}
\end{figure}
  
To reduce the mutual background, we study $\gamma_{\rm OS} - \gamma_{\rm SS}$, and multiply it by the number of participants, $N_{\rm part}$, to compensate for the dilution effect~\cite{STAR_LPV2}.
The right panel of Fig.~\ref{fig5} shows the signal $(\gamma_{\rm OS} - \gamma_{\rm SS})\cdot N_{\rm part}$  vs $v_{2}$ for different centralities in 193 GeV U+U and 200 GeV Au+Au collisions.
In both U+U and Au+Au, the signal roughly increases with $v_{2}$ pointing to the large background effect due to flow. The central trigger in U+U is supposed to disentangle the background contribution from the signal, since the magnetic field will be greatly suppressed and the measured signal will be dominated by the $v_2$-related background. As a result, in $0$-$1\%$ most central U+U collisions the signal disappears as expected by the Chiral Magnetic Effect, while $v_2$ is still $\sim 2.5\%$. 
                         
\section {Summary}
Azimuthal  anisotropy $v_{n}$ measurement for \textit {n} = 1-5 as a function of transverse momentum $p_{T}$ and centrality in U+U collisions at $\sqrt{s_{NN}}$ = 193 GeV is presented. We observe a weak centrality dependence for harmonics other than the second harmonic.  For higher harmonics and mid central collisions, $v_{n}$(U+U) is similar to $v_{n}$(Au+Au),  the difference appears at central collisions for $v_{1}$ and $v_{2}$.  In ultra central collisions   $v_{2}(U+U)$ shows a stronger multiplicity dependence compared to $v_{2}( Au+Au)$.  The signal related to LPV seems to disappear when the magnetic field is greatly suppressed as in 0-1\% most central UU collisions, while $v_{2}$ is still sizable indicating that observed signal is not dominated by the $v_{2}$-related background.

\end{document}